\documentclass[aip,jap,amsmath,amssymb,
reprint,
]{revtex4-1}

\usepackage{graphicx}
\usepackage{dcolumn}
\usepackage{bm}

\usepackage[mathlines]{lineno}

\begin{document}

\title[]{Conductance fluctuations in metallic nanogaps made by electromigration}

\author{P. Petit}
\author{A. Anthore}
\author{M.L. Della Rocca}
\email[Corresponding author~: ]{maria-luisa.della-rocca@univ-paris-diderot.fr}
\author{P. Lafarge}
\affiliation{Laboratoire Mat\'eriaux et Ph\'enom\`enes Quantiques, Universit\'e Paris Diderot-Paris 7, CNRS UMR 7162,
75205 Paris Cedex 13, France}

\date{\today}

\begin{abstract}
We report on low temperature conductance measurements of gold nanogaps fabricated by controlled electromigration. Fluctuations of the conductance due to quantum interferences and depending both on bias voltage and magnetic field are observed. By analyzing the voltage and magnetoconductance correlation functions we determine the type of electron trajectories generating the observed quantum interferences and the effective characteristic time of phase coherence in our device.
\end{abstract}

\maketitle

\section{INTRODUCTION }

Metallic nanogaps made by electromigration are key components of three terminal single molecule electronic devices. Besides their potential use as nanometer-spaced contact electrodes in future molecular devices, they are also ideal systems to investigate quantum transport properties at the nanoscale. As in other mesoscopic systems, quantum interferences are thus expected to have a strong influence on electronic transport properties of nanogaps at low temperature. The typical signatures of these phenomena in diffusive wire are weak localization and universal conductance fluctuations\cite{Montambaux, Nazarov1, Washburn}. Voltage dependent conductance fluctuations have been reported so far in related systems such as wires interrupted by a tunnel junction\cite{Mooij1,Mooij2,Mooij3,Nazarov2} or an atomic contacts\cite{VanRuitenbeek1,VanRuitenbeek2,Buhrman}, however they have not been investigated in metallic nanogaps.
This is an important issue for single molecule based transistors since the measurement and the operation of this kind of molecular devices can be hindered by quantum interference effects taking place in the wires, which produces variations of the current unrelated to the specific signatures of conduction processes through the single molecule.

In this article we report conductance fluctuations in gold electromigrated nanogaps as a function of voltage and magnetic field. By analyzing the correlation functions of the fluctuations we determine the physical mechanism responsible of quantum interferences and we extract the effective characteristic time of phase coherence in our device.

\section{EXPERIMENT }
\begin{figure}
\includegraphics{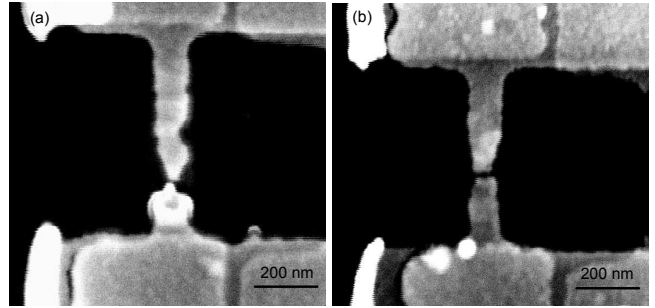}
\caption{(Color online) Scanning electron microscope images of sample 1 (a) and 2 (b) after electromigration and thermal cycling.}
\end{figure}

Samples are fabricated at room temperature in high vacuum (10$^{-6}$ mbar) by controlled electromigration of a gold nanowire 500 nm long, 100 nm wide and 20 nm thick. The electromigration procedure has been previously described\cite{Mangin1, Mangin2}, however the final steps leading to the nanogap formation have been changed following O'Neill et al.\cite{ONeill}. We initially thin the nanowire until the conductance reaches typical values of 3-4 $G{_0}$, $G_0=2e^2/h$ being the quantum of conductance for degenerate spins. Then we take advantage of the high diffusion constant of gold atoms at room temperature\cite{Gobel} by letting the nanowire spontaneously evolve while monitoring its conductance value. Atoms rearrangement is accompanied by conductance steps, indicating the evolution of the contact through a small number of conduction channels. In several hours a typical conductance value of 0.1 $G{_0}$ is obtained corresponding to bare nanogaps in the tunneling regime. In these conditions we cool down the sample to our dilution refrigerator base temperature (60 mK). The cooling down procedure typically decreases the conductance by one order of magnitude to 0.01-0.05 $G{_0}$.

We present results on two samples obtained with the same procedure but with different final conductance mean values, 0.0086 $G{_0}$ for sample 1 and 0.050 $G{_0}$ for sample 2. 
Scanning electron microscope (SEM) images of the two samples are shown on Fig. 1 (a) and (b) respectively. Starting from the 500 nm long nanowire, electromigration induces an asymmetric breaking. We note that for sample 1 the resulting electrodes lengths are $\sim $100 nm and $\sim $400 nm (Fig. 1 (a)), and for sample 2 we measure similarly $\sim $200 nm and $\sim $300 nm (Fig. 1 (b)). Such asymmetry in electrodes length plays a fundamental role in electron coherence as we will show in this article. Near-identical results for the conductance fluctuations as a function of magnetic field and bias voltage
were also found on a third broken nanowire fabricated on the same substrate, showing the same assymetry and having a conductance 0.0005 $G{_0}$. Note that SEM images are taken after warming up the samples, as a consequence the nanogaps sizes visible on Fig. 1 are enlarged due to thermal cycling.

Conductance measurements are performed using a lock-in amplifier technique with an AC voltage $V_{m}=30\mu$V at a frequency of 77 Hz. 
Electrical leads are filtered by an RC distributed line thermally anchored to each stage of the dilution unit, connecting the dilution refrigerator mixing chamber to room temperature. In particular a 1 m long RC distributed filter is located at the base temperature stage where the sample is mounted in a microwave-tight copper box.  The sample is positioned at the center of a superconducting magnetic coil, which allows to apply a perpendicular magnetic field.

\section{RESULTS}

Transport measurements on the examined samples show no signatures of Coulomb blockade or Kondo physics
%artifacts 
due to residual gold clusters ensuring that nanogaps are in a well established tunneling regime.
When measuring the samples the voltage drop is localized at the nanogap, since its conductance is much smaller than the conductance of the leads ($\sim$75 $G_{0}$). This allows to vary the energy interval over which electronic interferences occur resulting in voltage dependent conductance fluctuations.

Figs. 2 (a) and (b) show conductance measurements for sample 1 and 2 as a function of the bias voltage for different values of the magnetic field, from 0 to 1.3 T. A non linear background due to the voltage dependence of tunnel transport has been subtracted. Conductance fluctuations are clearly evident and reproducible. 
Note the presence of a well defined dip at V = 0 for all values of the magnetic field. 
This zero-bias anomaly (ZBA), independent of the sample, can be explained by dynamical Coulomb blockade due to the dissipative electromagnetic environment surrounding the nanogap.
In the inset of Fig. 2 (b) we analyse quantitatively this anomaly for the $G(V )$ data normalized to the tunnel conductance $G_T$ at B = 1.1 T. The dissipative environment of the nanogap is modeled as a parallel capacitance and a series resistance \cite{Joyez}. For an uncontrolled environment, the series resistance is expected to be close to the vacuum impedance Z$_{0}=\sqrt{\mu _0/\varepsilon _0}$ at microwave frequencies. The fit, shown as solid line, is then performed using two free parameters, the junction capacitance $C$ and the electronic temperature T$_e$, and fixing the series resistance R = Z$_0$=377 $\Omega$. The resulting value C = 10 $\pm$ 6~fF can be related by geometrical arguments to the capacitance towards the gate electrode. The found electronic temperature T$_e$ = 500 $\pm$ 100 mK is consistent with the temperature corresponding to the AC excitation voltage $\sim$ 360 mK, such strong excitation was needed to get accurate measurements. The uncertainties are quite high due to the difficulty to infer G$_T$ accurately because of the conductance fluctuations.

\begin{figure}
\includegraphics{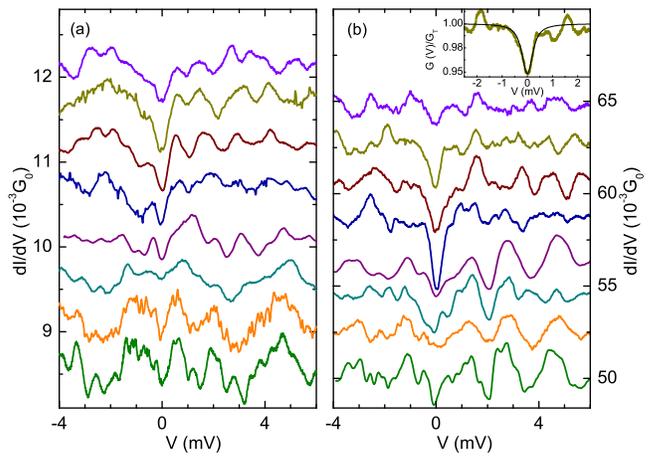}
\caption{(Color online) Conductance of sample 1 (a) and 2 (b) as a function of the bias voltage in the range -4 mV $\leq V \leq$6 mV for different values of the magnetic field (from bottom to top $B$=0, 0.1, 0.3, 0.5, 0.7, 0.9, 1.1, 1.3 T). Curves are vertically shifted for clarity. Inset: Dynamical Coulomb blockade fit\cite{Joyez} of the zero bias anomaly of the $G(V)$ curve at $B$= 1.1 T.}
\end{figure} 

\begin{figure}
\includegraphics{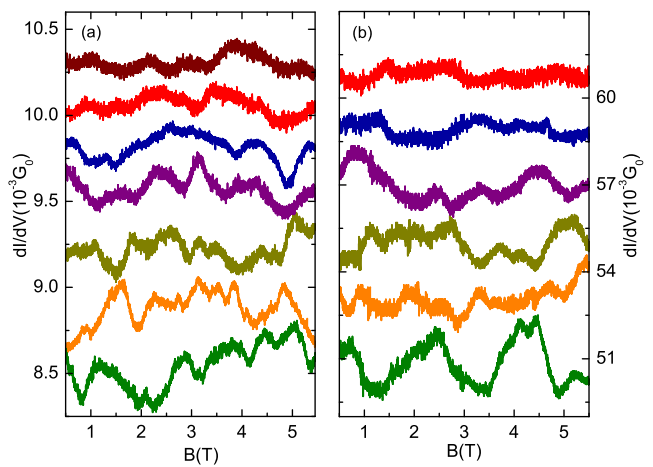}
\caption{(Color online) Conductance of sample 1 (a) and 2 (b) as a function of the magnetic field in the range 0.5 T $\leq B\leq$5.5 T for different values of the bias voltage (from bottom to top, $V$ = 1.1, 2.7, 4.1, 6.1, 7.1, 8.1, 10.1 mV for sample 1 and $V$ = 1.1, 3.1, 5.1, 7.1, 9.1, 11.1 for sample 2. Curves are vertically shifted for clarity.}
\end{figure}

When measuring the conductance as a function of the magnetic field at different bias voltages analogous fluctuations occur. Figs. 3 (a) and (b) show the $G(B)$ curves for different values of the bias voltage, from 1 to 11 mV, for sample 1 and 2 respectively. A linear background, where present, has been subtracted. Note that we have verified that the G(B) curves are symmetric with respect to B = 0.

\section{DISCUSSION}

The observed fluctuations are signatures of interference of electronic trajectories in metallic wires, which highlights the wave nature of electrons \cite{Washburn}. We analyze our data by calculating the correlation function normalized to the variance defined as $F(\Delta B, \Delta V)=\langle\delta G(V,B) \delta G(V+\Delta V, B+\Delta B)\rangle/\langle\delta G(V,B)^{2}\rangle$. In Fig. 4 (a) and (b) $F(\Delta B=0, \Delta V)$ is plotted as function of $\Delta V$ for the two samples (ZBA has not been taken into account in data computation). The experimental data (dots) are obtained from measurements reported in Fig. 2 (a) and (b), consisting of a set of 1000 data points. Following the ergodic hypothesis, changing the magnetic field corresponds to changing the impurity configurations in the wire, so data are averaged on all the explored magnetic fields values. The inset of Fig. 4 (a) and (b) shows $F(\Delta B, \Delta V=0)$ as function of $\Delta B$ for the two samples at V = 1.1 mV, obtained from a set of 2500 data points in the interval 0.5$\leq$B$\leq$3 T. 

These data are compared with two theoretical models, previously used for describing fluctuations in metallic wires interrupted by a tunnel junction \cite{Mooij1} and in atomic-size point contacts \cite{VanRuitenbeek2}. 
On the one hand, our device may behave as an atomic contact with low transmission coefficient since the spontaneous breaking of the nanowire occurs via discrete conductance steps indicating a small number of conduction channels.
On the other hand, during the cool down procedure we measure a further conductance decrease without characteristics steps. This reveals atomic rearrangement which may lead to a large number of low transmitted conduction channels, a configuration better described by the tunnel junction model.

In these models the energy of the injected electrons can be tuned by the applied bias voltage. 
Impurities in the leads allow diffusion of electrons through multiple scattering processes, so that once an electron is injected in the lead, it can be backscattered. In the tunnel junction model, conductance fluctuations are due to closed loop trajectories involving the two leads and including the junction. 
In the case of an atomic contact, conductance fluctuations are generated by interferences between electron waves directly transmitted by the contact and the waves backscattered by the diffusive leads and reflected at the contact towards the same lead. In both cases interference loops have in first approximation typical area equal to $L\times W$, where $W$ is the width of the electrodes and $L$ is the characteristic length limiting electron phase coherence. 
A change $\Delta B$ in the magnetic field applied perpendicularly to the sample adds a phase shift $\Delta \varphi =2\pi\Delta B L W/\phi_{0}$ to the electron wavefunction, where $\phi_{0}=h/e$ is the flux quantum. The typical field scale of fluctuations is given by $\phi_{0}/(L W)$. In the same way, a phase shift can be induced by an energy change $e\Delta V$ giving $\Delta \varphi = e\Delta V L^2/(\hbar D)$, where $D$ is the diffusion constant of electrons in gold. The corresponding voltage scale of fluctuations is $h D/(e L^2)$. 
The full theoretical expression for the correlation function in the two cases \cite{Mooij1,VanRuitenbeek2} is given respectively by
\begin{widetext}
\begin{equation}
\langle\delta G(V,B) \delta G(V+\Delta V, B+\Delta B)\rangle_{1}=\frac{4 V G_{T}^{2}}{\pi e \hbar (\rho V_{m})^{2}}
\int_ 0^\infty {P_{cl,1}^{2}(t) \cos(\frac{e}{\hbar}\Delta V t) e^{-2t/\tau }e^{-t/\tau_{B} }J_{1}^{2}(\frac{e}{\hbar} V_{m} t) \, dt}
\end{equation}
\begin{equation}
\langle\delta G(V,B) \delta G(V+\Delta V, B+\Delta B)\rangle_{2}=(\frac{2 e}{h})^{2} \frac{16 T^{2}(1-T)}{V_{m}^{2}}
\int_ 0^\infty {(\frac{\hbar}{\tau })^{2} P_{cl,2}(t) \cos(\frac{e}{\hbar}\Delta V t) e^{-t/\tau }e^{-t/\tau_{B} }J_{1}^{2}(\frac{e}{\hbar} V_{m} t) \, dt}
\end{equation}
\end{widetext}
where the index 1 and 2 refer to the tunnel junction and the atomic contact model. Here $\rho $ is the electrode density of states, $T$ the atomic contact transmission coefficient, $V$ the DC bias voltage, $\tau _{B}$ the dephasing time due to the magnetic field, given by $\tau _{B}=12 \frac{(\hbar /e \Delta B )^{2}}{2DW^{2}}$, and $\tau$ is the time limiting electron coherence. The first order Bessel function $J_{1}(\frac{e}{\hbar} V_{m} t)$ takes into account the dephasing due to the amplitude of the applied AC voltage $V_{m}$ which has to be considered if $e V_{m}>\hbar /\tau $. The function $P_{cl,i}(t)$ is the classical probability for an electron tunneling through the nanogap to be backscattered towards the nanogap again for the two models ($i = 1,2$). In the geometry and energy range considered here, transport is quasi one-dimensional thus $P_{cl,i}(t)=\frac{K_{i}}{\sqrt{\pi D t}}$, where $K_{i}$ is a factor including constant terms and geometrical details. 
The return probabilities appear with different power laws in Eq. (1) and (2). This is related to the  particular electronic trajectories generating quantum interferences considered in the two models.
In the tunnel junction model, the transport is supposed to be diffusive over the whole length of the wire except at the tunnel junction. Therefore, the diffusive trajectories contributing to conductance fluctuations comparable with the junction conductance, are closed loops crossing the tunnel barrier and involving both the left and right electrodes. As a consequence, assuming the diffusion constants on both sides are equal, Eq. (1) presents a square dependence on the classical return probability $P_{cl,1}(t)$. In the atomic contact model, the system is described as a central ballistic region containing the atomic contact surrounded by diffusive regions on either side.
The linear dependence on the return probability in Eq. (2) can be obtained by considering electron waves being transmitted by the contact and injected from the central ballistic part in one of the diffusive regions. They can be backscattered with probability $P_{cl,2}(t)$, reflected by the contact, and interfere with the directly transmitted waves, thus generating conductance fluctuations.

\begin{figure}
\includegraphics{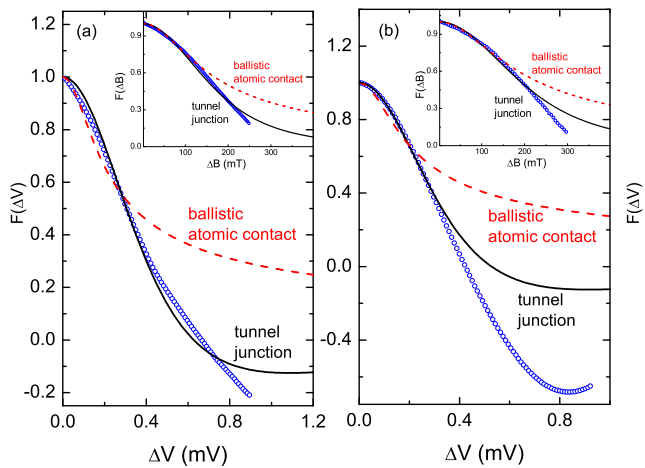}
\caption{(Color online) Normalized correlation function (blue dots) for sample 1 (a) and 2 (b) as a function of $\Delta V$. Insets: normalized correlation functions as a function of $\Delta B$. Data are compared with the tunnel junction model (black curve) and the atomic contact model (red dashed curve).}
\end{figure}

Full and dashed lines in Fig. 4 represent the fit to the experimental data on the basis of the two models. When fitting the voltage correlation function we set $W=10^{-7}$ m and $S=2\times $10$^{-15}$ m$^{2}$ according to the geometry of the samples. 
Moreover, since, at $\Delta B$ = 0, the calculated voltage correlation functions weakly depend on the diffusion constant, we take $D=0.005$ m$^{2}$/s estimated from the Einstein relation\cite{Montambaux} by measuring the nanowire resistivity before electromigration at room temperature\cite{note1}. The only free parameter is then the time $\tau $. Clearly the tunnel junction model seems more appropriate to describe our experimental data. 
This result indicates that even if during the fabrication process, the nanogap evolves through a small number of conduction channels, final atoms rearrangement at low temperature corresponds better to the tunnel junction limit of a large number of conduction channels with low transmission. Moreover the tunnel junction model suggests that interferences in our devices are due to trajectories involving the two electrodes and the nanogap.
Using Eq. (1) we extract $\tau $ obtaining similar results for the two samples, $\tau \sim $ 2.3$\pm$0.3 ps. Unlike the voltage correlation functions, the calculation of the magnetoconductance correlation functions at $\Delta V$ = 0 strongly depends on $D$ through the exponential dependence in $\tau _{B}$. In order to fit the experimental data (inset of Fig. 4 (a) and (b)), we fix $\tau $ at the value extracted from the former fit and we leave $D$ as free parameter. We obtain $D$ = 0.006$\pm$0.002 m$^{2}$/s in agreement with our previous estimate. Knowing $\tau $ we can calculate the length over which electron coherence is preserved by the formula $L=\sqrt{D \tau }$, giving $L\sim$ 110 nm. 
This result is consistent with the characteristic length extracted by estimating $\Delta V$ from data in Fig. 2 (a) and (b) using the formula $\Delta V=h D/(e L^{2})$. We obtain $L\approx$ 100-140 nm with $\Delta V\approx$ 1-2 mV.
These values are lower than the expected phase coherence time $\tau _{\phi }$ and length $L_{\phi }$ in gold, of the order of 200-300 ps and 20-30 $\mu$m respectively at the experimental temperature. 
This is not surprising by considering the typical dimensions of our devices. Electron coherence in fact is limited by the length of the nanogap electrodes and the large reservoirs. As shown in Fig 1, initially the nanowire is 500 nm long, after electromigration the nanogap is formed not in the center of the wire, the shorter electrode has length as small as 100 nm.
Note also that $L$ remains smaller than the thermal length $L_{T}=\sqrt{\hbar D/k_{B}T_{e}}$ = 280 nm for $T_{e}$ = 500 mK, indicating that electron coherence is not temperature limited.

\section{CONCLUSION}

In conclusion, we have measured conductance fluctuations as a function of the bias voltage and the magnetic field in gold electromigrated nanogaps at low temperature.
By analyzing the correlation functions on the basis of existing theoretical models, we conclude that the nanogap can not be described as a low transmitted atomic contact but rather as a tunnel element interrupting a diffusive wire. Fluctuations are due to quantum interferences related to closed electronic trajectories limited by sample geometry and including the nanogap. 
%Increasing the length of the wire used to form the nanogap would allow to probe the intrinsic %electronic coherence length
Since, in our device, electron coherence is limited by the length of the nanogap electrodes, increasing the length of the wire used to form the nanogap would increase the effective characteristic length of phase coherence and eventually would allow to probe the intrinsic electronic coherence length. 
This study shows 
that electron coherence in contact electrodes has to be taken into account in future single molecule electronic devices.
Especially the effect of unwanted conductance fluctuations may be minimized by using short and wide source/drain electrodes in the design of three terminal single molecule devices.

\begin{acknowledgments}
This work was supported by the C$'$Nano Ile de France through the SPINMOL contract.
We gratefully acknowledge helpful discussions with the Quantronics group, T. Kontos and F. Pierre.
\end{acknowledgments}


\begin{thebibliography} {2}
\bibitem{note1} We have checked that the nanowire resistance does not change at cryogenics temperature, condition in which transport measurements are realized.
\end{thebibliography}
\end{document}